\documentclass[aps,showpacs,preprintnumbers,amsmath,amssymb,nofootinbib]{revtex4}
\usepackage{amssymb}
\usepackage{amsfonts}
\usepackage{amsmath}
\usepackage{graphicx}
\usepackage{dcolumn}
\usepackage{bm}
\usepackage[latin1]{inputenc}
\usepackage[dvips]{hyperref}

\usepackage{graphicx}
\def\be{\begin{equation}}
 \def\ee{\end{equation}}
 \def\bea{\begin{eqnarray}}
 \def\eea{\end{eqnarray}}

\begin{document}

\title{Fermionic Greybody Factors of Two and Five-Dimensional Dilatonic Black Holes}
\author{Ram\'{o}n B\'{e}car}
\email{rbecar@uct.cl}
\affiliation{Departamento de Ciencias Matem\'{a}ticas y F\'{\i}sicas, Universidad Cat\'{o}%
lica de Temuco, Montt 56, Casilla 15-D, Temuco, Chile}
\author{P. A. Gonz\'{a}lez}
\email{pablo.gonzalez@udp.cl}
\affiliation{Facultad de Ingenier\'{\i}a, Universidad Diego Portales, Avenida Ej\'{e}%
rcito Libertador 441, Casilla 298-V, Santiago, Chile.}
\author{Yerko V\'{a}squez.}
\email{yvasquez@userena.cl}
\affiliation{Departamento de F\'{\i}sica, Facultad de Ciencias, Universidad de La Serena,\\ 
Avenida Cisternas 1200, La Serena, Chile.}
\date{\today }

\begin{abstract}

We study fermionic perturbations in the background of a two and five-dimensional dilatonic black holes.
Then, we compute the reflection and transmission coefficients and the absorption cross section for fermionic fields, and we show numerically that the absorption cross section vanishes in the low and high frequency limit. 
Also we find that beyond a certain value of the horizon radius $r_0$ the absorption cross section for five-dimensional dilatonic black hole is constant. Besides, we have find  that the absorption cross section decreases for higher angular momentum, and it decreases when the mass of the fermionic field increases.

\end{abstract}

\maketitle


\section{Introduction}

The two-dimensional models of gravity are locally trivial and it is necessary to incorporate extra fields to add richness to the gravity model due to the two-dimensional Einstein-Hilbert action is just a topological inviariant (Gauss-Bonnet term). In this sense, the dilatonic field plays the role of the extra fields, which naturally arises, for instance, in the compactifications from higher dimensions or from string theory. These theories also have black hole solutions which play an important role in revealing various aspects of the geometry of spacetime and quantization of gravity, and also the physics related to string theory \cite{Witten:1991yr, Teo:1998kp, McGuigan:1991qp}. On the other hand, two-dimensional low-energy string theory admits several black hole solutions. Furthermore, technical simplifications in two dimensions often lead to exact results, and it is hoped that this might helps to address some of the conceptual problems posed by quantum gravity in higher dimensions. The exact solvability of two-dimensional models of gravity have proven to be a useful tool for investigations into black hole thermodynamics \cite{Lemos:1996bq,Youm:1999xn, Davis:2004xb, Grumiller:2007ju, Quevedo:2009ei, Belhaj:2013vza}. Such investigations are hoped to provide a deeper understanding of key issues; including the microscopic origin of black hole entropy \cite{Myers:1994sg, Sadeghi:2007kn, Hyun:2007ii}, and the end point of black hole evaporation via thermal radiation \cite{Kim:1999ig, Vagenas:2001sm, Easson:2002tg}. For an excellent review about dilaton gravity in two dimensions see \cite{Grumiller:2002nm}.

On the other hand, there is a growing interest in five-dimensional dilatonic black holes in the last few years,
since it is believed that these black holes can shed some light into the solution of the fundamental problem of the
microscopic origin of the Bekenstein-Hawking entropy. The area-entropy relation $S_{BH} = A/4$ was obtained for a class of five-dimensional extremal black holes in Type II string theory using D-brane techniques \cite{Strominger:1996sh}. In \cite{Teo:1998kp}, the author derived the entropy for the two-dimensional black hole \cite{McGuigan:1991qp} by establishing the U-duality between the two-dimensional black hole and the five-dimensional one \cite{Teo:1998kp}. A similar work was carried out in \cite{LopesCardoso:1998sq} using a different sequence duality transformations, this time in four dimensions and leading to the same expressions for the entropy for two-dimensional black holes. Besides, the issue about classical and quantum stability of two dimensional and five dimensional dilatonic black holes was carried out, for instance in \cite{Kim:1994nq, Nojiri:1998yg, Becar:2007hu, Becar:2010zz, LopezOrtega:2009zx, LopezOrtega:2011sc}.

Additionally, several studies          have contributed to the scattering and absorption properties of waves in the spacetime of black holes. As the geometry of the spacetime surrounding a black hole is non-trivial, the Hawking radiation emitted at the event horizon may be modified by this geometry, so that when an observer located very far away from the black hole measures the spectrum, this will no longer be that of a black body \cite{Maldacena:1996ix}. The factors that modify the spectrum emitted by a black hole 
are known as greybody factors and can be obtained through the classsical scattering; therefore its study allows to increase the semiclassical gravity dictionary, and also permits to gain insight into the quantum nature of black holes and, thus, of quantum gravity, for an excellent review about this topic see \cite{Harmark:2007jy}. Also, see for instance \cite{Moderski:2008nq, Gibbons:2008gg, Rogatko:2009jp}, for decay of Dirac fields in higher dimensional black holes. In the present work, the reflection and the transmission coefficients, and the greybody factors of two-dimensional stringy black holes (\cite{Witten:1991yr}, \cite{Frolov:2000jh}) and five-dimensional black holes \cite{Teo:1998kp} for fermionic fields are computed.

This paper is organized as follows. In Sec. \ref{FP}, we study fermionic perturbations in the background of two-dimensional dilatonic black holes, and in Sec. \ref{coeff} we calculate the reflection and the transmission coefficients, and the absorption cross section. Then, in Sec. \ref{FP2} and \ref{coeff2} we extend our previous results to the five-dimensional dilatonic black holes. Finally, our conclusions are in Sec. \ref{remarks}.

\section{Fermionic perturbations of two-dimensional dilatonic black holes}
\label{FP}
In order to have a gravity theory with dynamical degrees of
freedom in two-dimensional spacetime, we consider the gravity coupled to a
dilatonic field described by the action
\begin{equation}
S_{g}=\frac{1}{2\pi }\int d^{2}x\sqrt{-g}e^{-2\phi }\left( R+4(\nabla \phi
)^{2}+4\lambda ^{2}\right)~.  \label{accion}
\end{equation}
The field equations for the metric and dilaton are given by
\begin{eqnarray}
\beta _{\mu \nu }^{G} &=&R_{\mu \nu }+2\nabla_\mu \nabla_\nu \phi
=0~,\label{betag} \\
\beta ^{\phi } &=& \Box \phi -2\left( \nabla
\phi \right) ^{2}+2\lambda ^{2}=0~. \label{betaphi}
\end{eqnarray}
A general static metric describing a black hole in this theory can be written as
\begin{equation}
ds^{2}=-f(r)d\tau ^{2}+\frac{dr^{2}}{f(r)}~,  \label{metrica1}
\end{equation}
where $f(r)=1-e^{-\phi }$ and $\phi =(r-r_{0})/r_{0}$. The change
of coordinate $x=\frac{r-r_{0}}{r_{0}}$, yields $f(x)=1-e^{-x}$ with the horizon of the black hole located at
$x=0$. This solution represents a well-known string-theoretic black hole \cite{Teo:1998kp,McGuigan:1991qp,Witten:1991yr,Frolov:2000jh}. 
The fermionic perturbations in the background of two-dimensional dilatonic black holes are governed by the Dirac
equation
\begin{equation}\label{DE}
\left( \gamma ^{\mu }\nabla _{\mu }+m\right) \psi =0~,
\end{equation}%
where the covariant derivative is defined as 
\begin{equation}
\nabla _{\mu }=\partial _{\mu }+\frac{1}{2}\omega _{\text{ \ \ \ }\mu
}^{ab}J_{ab}~,
\end{equation}%
and the generators of the Lorentz group $J_{ab}$ are 
\begin{equation}
J_{ab}=\frac{1}{4}\left[ \gamma _{a},\gamma _{b}\right] ~.
\end{equation}%
The gamma matrices in curved spacetime $\gamma ^{\mu }$ are defined by 
\begin{equation}
\gamma ^{\mu }=e_{\text{ \ }a}^{\mu }\gamma ^{a}~,
\end{equation}
where $\gamma ^{a}$ are the gamma matrices in flat spacetime. In order to
solve the Dirac equation we use the diagonal vielbein 
\begin{equation}
e^{0}=\sqrt{f\left( r\right)} dt~,\text{ \ }e^{1}=\frac{1}{\sqrt{f\left( r\right) }}dr~.
\end{equation}
From the null torsion condition 
\begin{equation}\label{NT}
de^{a}+\omega _{\text{ \ }b}^{a}e^{b}=0~,
\end{equation}%
we obtain the spin connection 
\begin{equation}
\omega ^{01}=\frac{f^{\prime}\left( r\right)}{2\sqrt{f\left( r\right)}} e^{0},
\end{equation}
Now, by using
the following representation of the gamma matrices 
\begin{equation}
\gamma ^{0}=i\sigma ^{2} ~,\text{ \ }\gamma ^{1}=\sigma
^{1}~,
\end{equation}%
where $\sigma ^{i}$ are the Pauli matrices, 
along with the
following ansatz for the fermionic field 
\begin{equation}
\psi =e^{-i\omega t}\left( 
\begin{array}{c}
\psi _{1} \\ 
\psi _{2}%
\end{array}%
\right)~,
\end{equation}%
we obtain the
following equations 
\begin{eqnarray}\label{system1}
\sqrt{f}\partial _{r}\psi _{1}+\frac{f^{\prime}}{4\sqrt{f}}\psi _{1}+\frac{i\omega}{\sqrt{f}}\psi _{1}+m\psi _{2}&=&0~\notag \\
\sqrt{f}\partial _{r}\psi _{2}+\frac{f^{\prime}}{4\sqrt{f}}\psi _{2}-\frac{i\omega}{\sqrt{f}}\psi _{2}+m\psi _{1}&=&0~.
\end{eqnarray}%
By decoupling the system of equations and using 
\begin{equation}
\psi _{1}^{\prime }=z^{\alpha }\left( 1-z\right) ^{\beta }F\left( z\right)~, 
\end{equation}
where
\begin{equation}
z = 1-e^{-\phi }~, 
\end{equation}
and
\begin{equation}
\alpha =-\left(\frac{1}{4}+i\omega r_0\right)~, 
\end{equation}
\begin{equation}
\beta =r_0\sqrt{m^2-\omega^2}~, 
\end{equation}
we obtain the following equation for $F\left(z\right)$
\begin{equation}\label{HE}
z\left( 1-z\right) F^{\prime \prime }\left( z\right) +\left( c-\left(
1+a+b\right) z\right) F^{\prime }\left( z\right) -abF\left( z\right) =0~, 
\end{equation}
whose solution is given by 
\begin{equation}
\psi _{1}=C_{1}z^{\alpha }\left( 1-z\right) ^{\beta }{_2F_1}\left(
a,b,c,z\right) +C_{2}z^{1/2-\alpha }\left( 1-z\right) ^{\beta }{_2F_1}\left(
a-c+1,b-c+1,2-c,z\right)~, 
\end{equation}
which has three regular singular points at $z=0$, $z=1$ and $z = \infty$. Here, $_2F_1(a,b,c;z)$ is a hypergeometric function and $C_1$, $C_2$ are constants, and
\begin{equation}
a=\frac{3}{4}+\alpha +\beta~, 
\end{equation}
\begin{equation}
b=\frac{1}{4}+\alpha +\beta~, 
\end{equation}
\begin{equation}
c=1+2\alpha~. 
\end{equation}
Now, imposing boundary conditions at the horizon, i.e., that there is only ingoing modes, implies that $C_{2}=0$. Thus,  the solution can be written as
\begin{equation}
\label{Rhorizon2}
\psi _{1}=C_1z^{\alpha }\left( 1-z\right) ^{\beta }{_2F_1}\left(
a,b,c,z\right)~, 
\end{equation}
and by using the integrating factor 
$e^{-\int \left(\frac{\frac{1}{4}-i\omega r_0}{z}+\frac{i \omega r_0}{1-z}\right)dz}$, in Eq. (\ref{system1}), we get the solution 
\begin{equation}
\psi _{2}=-C_1mr_0z^{-\frac{1}{4}+i\omega r_0 }\left(
1-z\right) ^{-\omega r_0}\int z^{\prime c-1}\left(
1-z^{\prime }\right) ^{a-c-1}{_2F_1}\left( a,b,c,z^{\prime }\right)
dz^{\prime }~.
\end{equation}
So, if we consider the relation
\begin{equation}
\int z^{c-1}\left( 1-z\right) ^{a-c-1}{_2F_1}\left( a,b,c,z\right) dz=\left(
1-z\right) ^{a-c}z^{c}\frac{{_2F_1}\left( a,b+1,c+1,z\right) }{c}~,
\end{equation}
$\psi _{2}$ can be rewritten as
\begin{equation}\label{psi2horizon2}
\psi _{2}=-\frac{C_1mr_0z^{\frac{1}{4}-i\omega r_0}(1-z)^{r_0\sqrt{m^2-\omega^2}}}{\frac{1}{2}-2i\omega r_0}
{_2F_1}\left( a,b+1,c+1,z\right)~. 
\end{equation}

\section{Reflection coefficient, transmission coefficient and absorption cross section of two-dimensional diatonic black hole}
\label{coeff}
\begin{figure}[h]
\begin{center}
\includegraphics[width=0.45\textwidth]{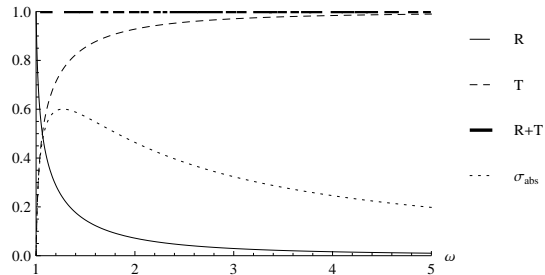}
\caption{The reflection coefficient $R$ (solid curve), the transmission coefficient $T$ (dashed curve),  $R+T$ (thick curve), and the absorption cross section $\sigma_{abs}$ (dotted curve) as a function of $\omega$, $(1\leq\omega)$; for $m=1$.}
\label{Coefficients1m1}
\end{center}
\end{figure}

The reflection and transmission coefficients depend on the behaviour
of the radial function both, at the horizon and at the asymptotic
infinity and they are defined by
\begin{equation}\label{reflectiond}\
R :=\left|\frac{\mathcal{F}_{\mbox{\tiny asymp}}^{\mbox{\tiny
out}}}{\mathcal{F}_{\mbox{\tiny asymp}}^{\mbox{\tiny in}}}\right|; \; T:=\left|\frac{\mathcal{F}_{\mbox{\tiny
hor}}^{\mbox{\tiny in}}}{\mathcal{F}_{\mbox{\tiny asymp}}^{\mbox{\tiny
in}}}\right|~,
\end{equation}
where $\mathcal{F}$ is the flux, and is given by 
\begin{equation}\label{flux}
\mathcal{F} =\sqrt{-g}\bar{\psi}\gamma ^{r}\psi~, 
\end{equation}
where, $\gamma ^{r}=e_{1}^{r}\gamma ^{1}$, $\bar{\psi}=\psi ^{\dagger
}\gamma ^{0}$,
$\sqrt{-g}=1$, 
and
$e_{1}^{r}=\sqrt{f\left(r\right)}$,
which yields
\begin{equation}\label{flux1} 
\mathcal{F}=\sqrt{f\left(r\right)}\left(
\left\vert \psi _{1}\right\vert ^{2}-\left\vert \psi _{2}\right\vert
^{2}\right)~.
\end{equation}
The behaviour at the horizon is given by~(\ref{Rhorizon2}), and using~(\ref{flux1}), we get the
flux at the horizon
\begin{equation}
\mathcal{F}
_{hor}^{in}= |C_{1}|^{2}~.
\end{equation}
To obtain the asymptotic behaviour of $\psi_1(r)$ and $\psi_2(r)$, we use $f(r)\rightarrow 1$, when $r\rightarrow \infty$ in~(\ref{system1}). Thus, we obtain the following solutions
\begin{equation}
\psi_{1,2}(r)=C_{1,2}e^{r\sqrt{m^2-\omega^2}}+D_{1,2}e^{-r\sqrt{m^2-\omega^2}}~.
\end{equation} 
Thus, the flux~(\ref{flux1}) at the asymptotic region is given by
\begin{equation}\label{fluxdinfinity}\
\mathcal{F}_{asymp}= \left|C_1\right|^2+\left|D_1\right|^2-\left|C_2\right|^2-\left|D_2\right|^2~.
\end{equation}
where $\omega^2 \geq m^2$. On the other hand,  by replacing the Kummer's formula \cite{M. Abramowitz}, in (\ref{Rhorizon2}) and (\ref{psi2horizon2}), 
\begin{eqnarray}\label{KE}
\nonumber {_2F_1}\left( a,b,c,z\right)  &=&\frac{\Gamma \left( c\right) \Gamma \left(
c-a-b\right) }{\Gamma \left( c-a\right) \Gamma \left( c-b\right) }%
{_2F_1}\left( a,b,a+b-c,1-z\right) + \\
&&\left( 1-z\right) ^{c-a-b}\frac{\Gamma \left( c\right) \Gamma \left(
a+b-c\right) }{\Gamma \left( a\right) \Gamma \left( b\right) }{_2F_1}\left(
c-a,c-b,c-a-b+1,1-z\right)~, 
\end{eqnarray}
and by using Eq. (\ref{flux1}) we obtain the flux
\begin{equation}\label{fluxdinfinity}\
F_{asymp}= \left|A_1\right|^2+\left|A_2\right|^2-\left|B_1\right|^2-\left|B_2\right|^2~.
\end{equation}
where,
\begin{eqnarray}
\nonumber A_1  &=&C_1\frac{\Gamma \left( c\right) \Gamma \left(
c-a-b\right) }{\Gamma \left( c-a\right) \Gamma \left( c-b\right) }~,\\
\nonumber A_2  &=&C_1\frac{\Gamma \left( c\right) \Gamma \left(a+b-c\right) }{\Gamma \left(a\right) \Gamma \left(b\right) }~,\\
\nonumber B_1  &=&-C_1mr_0\frac{\Gamma \left( c\right) \Gamma \left(c-a-b\right) }{\Gamma \left(c+1-a\right) \Gamma \left(c-b\right) }~,\\
B_2  &=&-C_1mr_0\frac{\Gamma \left( c\right) \Gamma \left(a+b-c\right) }{\Gamma \left(a\right) \Gamma \left(b+1\right) }~.
\end{eqnarray}
Therefore, the reflection and transmission coefficients
 are given by
\begin{equation}
R=\frac{|B_1|^{2}+|B_2|^{2}}{|A_1|^{2}+|A_2|^{2}}
~,\label{coef12}
\end{equation}
\begin{equation}
T=\frac{|C_1|^{2}}{|A_1|^{2}+|A_2|^{2}}
~,\label{coef22}
\end{equation}
and the absorption cross section $\sigma_{abs}$, becomes
\begin{equation}\label{absorptioncrosssection2}\
\sigma_{abs}=\frac{1}{\omega}\frac{|C_1|^{2}}{|A_1|^{2}+|A_2|^{2}}~.
\end{equation}
Now, we will carry out a  numerical analysis of the reflection coefficient~(\ref{coef12}), transmission coefficient~(\ref{coef22}), and absorption cross section~(\ref{absorptioncrosssection2}) of  two-dimensional dilatonic black holes, for fermionic fields.
So, we plot the reflection and transmission
coefficients and the absorption cross section in Fig.~(\ref{Coefficients1m1}), for fermionic fields with $m=1$. 
Essentially, we found that the reflection coefficient is one in the low frequency limit, that is $\omega\approx m$, and for high frequency limit this coefficient is null, being the behavior of the transmission coefficient opposite, with $R+T=1$. 
Also, the absorption cross section is null in the low and high-frequency limit, but there is a range of frequencies for which the absorption cross section is not null, and also it has a maximum value, see Fig.~(\ref{Coefficients1m1}). 
\section{Fermionic perturbations of five-dimensional dilatonic black holes}
\label{FP2}
The metric for five-dimensional dilatonic black holes can be written as the product of two completely decoupled parts, namely an asymptotically flat two-dimensional geometry that describes a two-dimensional  dilatonic  black hole (\ref{metrica1}) and a three-sphere with a constant radius $r_0$. The metric can be written as \cite{Teo:1998kp}
\begin{equation}
ds^2=-f(r)dt^2+\frac{1}{f(r)}dr^2+r_0^2{\rm
d}\Omega_3{}^2~.\label{5metric}
\end{equation} 
Now, in order to solve the Dirac equation (\ref{DE}), we use the diagonal vielbein 
\begin{equation}
e^{0}=\sqrt{f\left( r\right)} dt~,\text{ \ }e^{1}=\frac{1}{\sqrt{f\left( r\right) }}dr~, \text{ \ }e^m=r_0\tilde e^m~,
\end{equation}
and from the null torsion condition (\ref{NT}) 
the spin connection yields
\begin{equation}
\omega ^{01}=\frac{f^{\prime}\left( r\right)}{2\sqrt{f\left( r\right)}} e^{0}~, \text{ \ } \omega^{mn}=\tilde \omega^{mn}~, 
\end{equation}
where $\tilde \omega^{mn}$ is the connection of the base manifold. Also, by using
the following representation of the gamma matrices 
\begin{equation}
\gamma ^{0}=i\sigma ^{2}\otimes \mathbf{1}~,\text{ \ }\gamma ^{1}=\sigma
^{1}\otimes \mathbf{1}~,
\text{ \ }\gamma ^{m}=\sigma ^{3}\otimes \tilde{\gamma}^{m}~,
\end{equation}%
where $\sigma ^{i}$ are the Pauli matrices, 
and $\tilde{\gamma}^{m}$ are the
Dirac matrices in the base manifold $\Omega_3$,
along with the
following ansatz for the fermionic field 
\begin{equation}
\psi =e^{-i\omega t}\left( 
\begin{array}{c}
\psi _{1} \\ 
\psi _{2}%
\end{array}%
\right)\otimes \eta~,
\end{equation}%
we obtain the
following equations 
\begin{eqnarray}\label{system}
\sqrt{f}\partial _{r}\psi _{1}+\frac{f^{\prime}}{4\sqrt{f}}\psi _{1}+\frac{i\omega}{\sqrt{f}}\psi _{1}+(m-\frac{i\kappa}{r_0})\psi _{2}&=&0~\notag \\
\sqrt{f}\partial _{r}\psi _{2}+\frac{f^{\prime}}{4\sqrt{f}}\psi _{2}-\frac{i\omega}{\sqrt{f}}\psi _{2}+(m+\frac{i\kappa}{r_0})\psi _{1}&=&0~,
\end{eqnarray}%
where $i\kappa= \pm i(l+3/2)$ is the eigenvalue of the Dirac operator on the three-sphere. 
By decoupling the system of equations and using 
\begin{equation}
\psi _{1}^{\prime }=z^{\alpha }\left( 1-z\right) ^{\beta }F\left( z\right)~, 
\end{equation}
where
\begin{equation}
z = 1-e^{-\phi }~, 
\end{equation}
and
\begin{equation}
\alpha =-\left(\frac{1}{4}+i\omega r_0\right)~, 
\end{equation}
\begin{equation}
\beta =r_0\sqrt{m^2+\frac{\kappa^2}{r_0^2}-\omega^2}~, 
\end{equation}
we obtain (\ref{HE}),
whose solution is given by 
\begin{equation}
\psi _{1}=\bar{ C}_{1}z^{\alpha }\left( 1-z\right) ^{\beta }{_2F_1}\left(
a,b,c,z\right) +\bar{ C}_{2}z^{1/2-\alpha }\left( 1-z\right) ^{\beta }{_2F_1}\left(
a-c+1,b-c+1,2-c,z\right)~, 
\end{equation}
which has three regular singular points at $z=0$, $z=1$ and $z = \infty$. Here, $_2F_1(a,b,c;z)$ is a hypergeometric function and $\bar{ C}_1$, $\bar{ C}_2$ are constants, and
\begin{equation}
a=\frac{3}{4}+\alpha +\beta~, 
\end{equation}
\begin{equation}
b=\frac{1}{4}+\alpha +\beta~, 
\end{equation}
\begin{equation}
c=1+2\alpha~. 
\end{equation}
Now, imposing boundary conditions at the horizon, i.e., that there is only ingoing modes, implies that $\bar{ C}_{2}=0$; and in a similar way to  two-dimensional dilatonic black holes, we obtain the following solution for $\psi_1$ and $\psi_2$:
\begin{equation}
\label{Rhorizon}
\psi _{1}=\bar{ C}_1z^{\alpha }\left( 1-z\right) ^{\beta }{_2F_1}\left(
a,b,c,z\right)~, 
\end{equation}
\begin{equation}\label{psi2horizon}
\psi _{2}=-\frac{\bar{ C}_1(m+\frac{i\kappa}{r_0})r_0z^{\frac{1}{4}-i\omega r_0}(1-z)^{r_0\sqrt{m^2+\frac{\kappa^2}{r_0^2}-\omega^2}}}{\frac{1}{2}-2i\omega r_0}
{_2F_1}\left( a,b+1,c+1,z\right)~. 
\end{equation}

\section{Reflection coefficient, transmission coefficient and absorption cross section of five-dimensional dilatonic black holes}
\label{coeff2}
\begin{figure}[h]
\begin{center}
\includegraphics[width=0.45\textwidth]{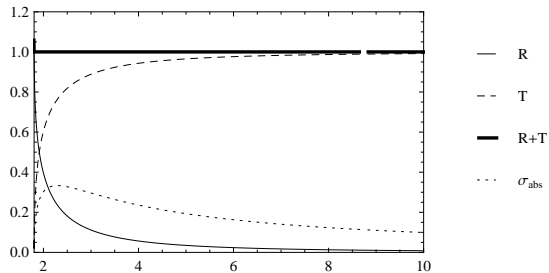}\\
\caption{The reflection coefficient $R$ (solid curve), the transmission coefficient $T$ (dashed curve),  $R+T$ (thick curve), and the absorption cross section $\sigma_{abs}$ (dotted curve) as a function of $\omega$, $(1.5\leq\omega)$; for $m=1$, $r_0=1$, and $l=0$.}
\label{Coeff5dk0}
\end{center}
\end{figure}

\begin{figure}[h]
\begin{center}
\includegraphics[width=0.45\textwidth]{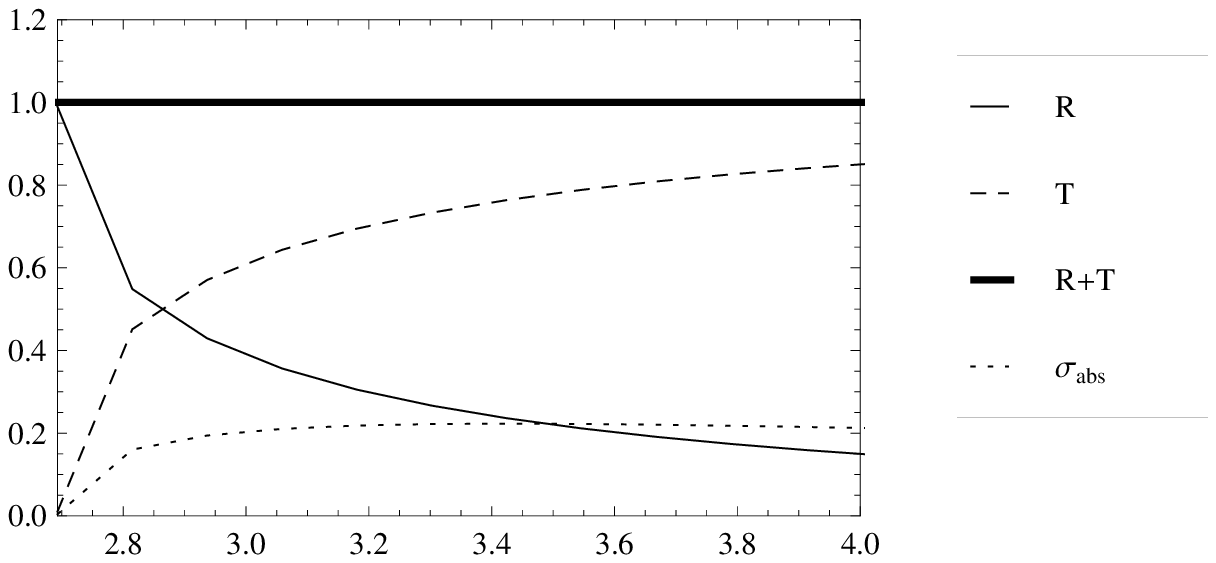}
\caption{The reflection coefficient $R$ (solid curve), the transmission coefficient $T$ (dashed curve),  $R+T$ (thick curve), and the absorption cross section $\sigma_{abs}$ (dotted curve) as a function of $\omega$, $(3.693\leq\omega)$; for $m=1$, $r_0=1$, and $l=1$.}
\label{Coeff5dk3}
\end{center}
\end{figure}

\begin{figure}[h]
\begin{center}
\includegraphics[width=0.45\textwidth]{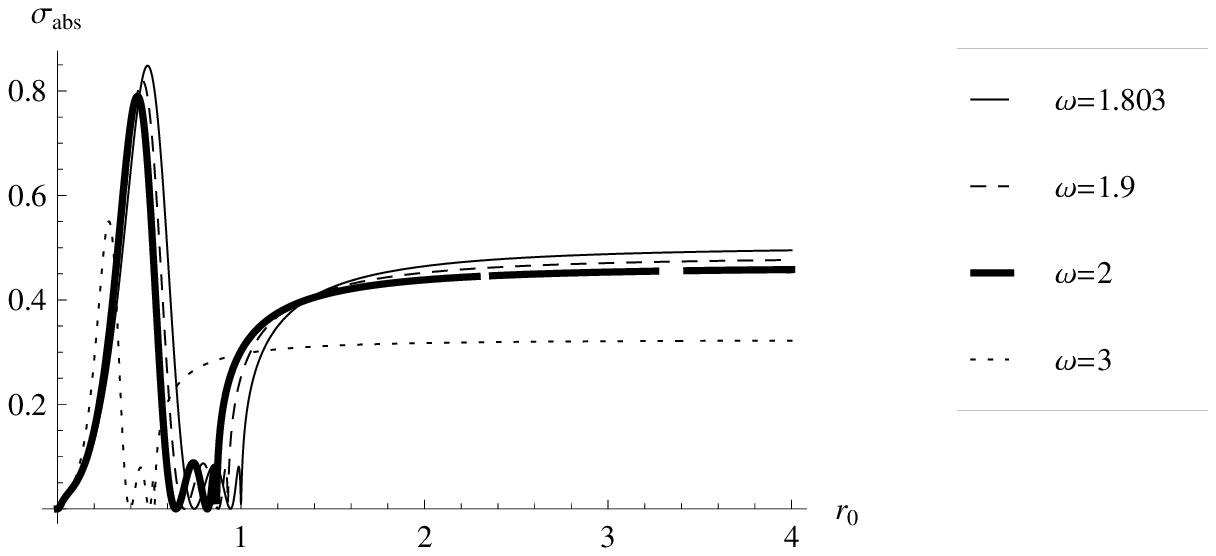}
\caption{The absorption cross section $\sigma_{abs}$ as a function of $r_0$, for $m=1$, $l=0$, $\omega=1.803$ (thin curve), $\omega=1.9$ (dashed curve), $\omega=2$ (thick curve), and $\omega=3$ (dotted curve).}
\label{ACSr0k0}
\end{center}
\end{figure}

\begin{figure}[h]
\begin{center}
\includegraphics[width=0.45\textwidth]{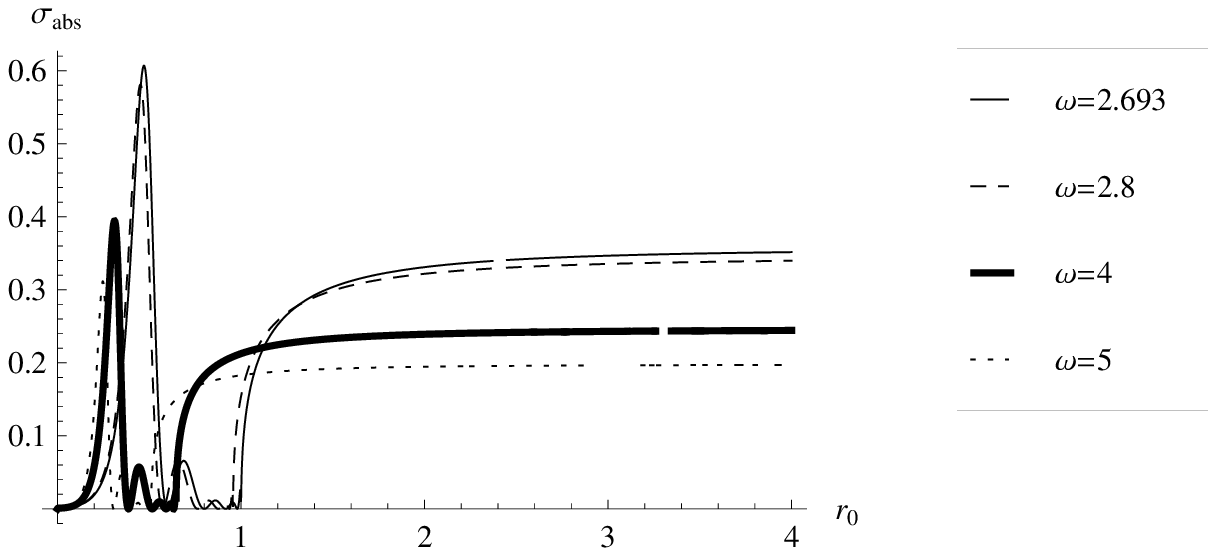}
\caption{The absorption cross section $\sigma_{abs}$ as a function of $r_0$, for $m=1$, $l=1$, $\omega=2.693$ (thin curve), $\omega=2.8$ (dashed curve), $\omega=4$ (thick curve), and $\omega=5$ (dotted curve).}
\label{ACSr0k3}
\end{center}
\end{figure}

\begin{figure}[h]
\begin{center}
\includegraphics[width=0.45\textwidth]{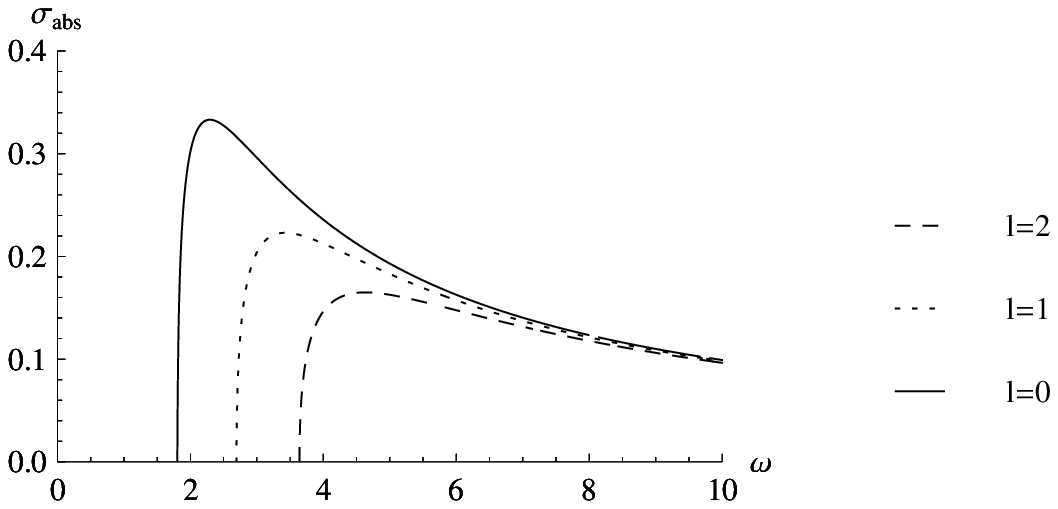}
\end{center}
\caption{The behaviour of $\sigma_{abs}$ as a function of $\omega$, for $m=1$, $r_0=1$ and $l=0,1,2$.} 
\label{Finall}
\end{figure}

\begin{figure}[h]
\begin{center}
\includegraphics[width=0.45\textwidth]{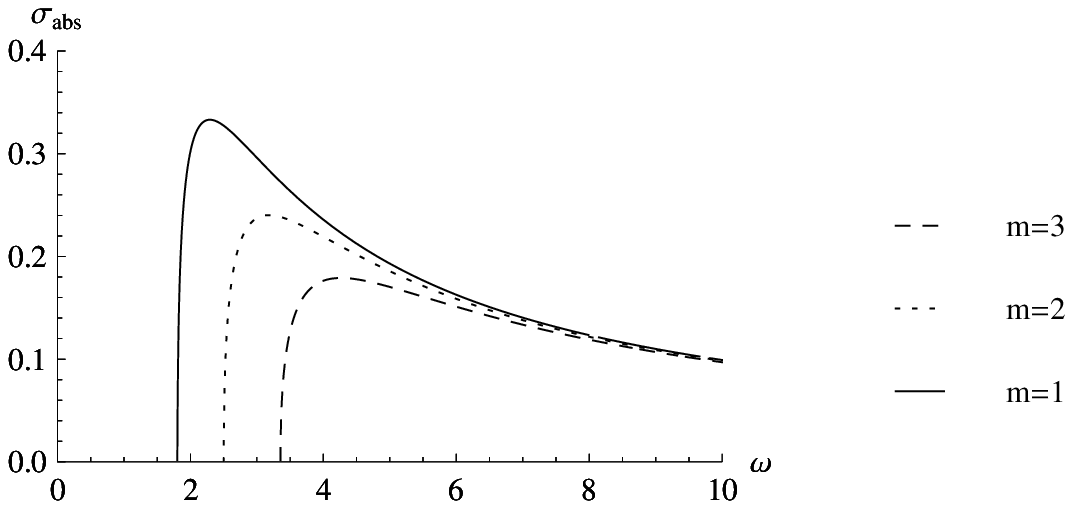}
\end{center}
\caption{The behaviour of $\sigma_{abs}$ as a function of $\omega$, for $l=0$, $r_0=1$ and $m=1,2,3$.} \label{Finalm}
\end{figure}

As we mentioned, the reflection and transmission coefficients depend on the behaviour
of the radial function both, at the horizon and at the asymptotic
infinity and they are defined by (\ref{reflectiond})
where $\mathcal{F}$ is the flux, and is given by (\ref{flux}), and yields  
\begin{equation}\label{flux2} 
\mathcal{F}\propto\sqrt{f\left(r\right)}\left(
\left\vert \psi _{1}\right\vert ^{2}-\left\vert \psi _{2}\right\vert
^{2}\right)~,
\end{equation}
where we have used $\gamma ^{r}=e_{1}^{r}\gamma ^{1}$, $\bar{\psi}=\psi ^{\dagger
}\gamma ^{0}$, 
and
$e_{1}^{r}=\sqrt{f\left(r\right)}$.
The behaviour at the horizon is given by~(\ref{Rhorizon}), and using~(\ref{flux2}), we get the
flux at the horizon up to an irrelevant factor from the angular
part of the solution
\begin{equation}
\mathcal{F}_{hor}^{in}= |\bar{ C}_{1}|^{2}~.
\end{equation}
Now, in order to obtain the asymptotic behaviour of $\psi_1(r)$ and $\psi_2(r)$, we use $f(r)\rightarrow 1$, when $r\rightarrow \infty$ in~(\ref{system}). Thus, we obtain the following solutions
\begin{equation}
\psi_{1,2}(r)=\bar{ C}_{1,2}e^{r\sqrt{m^2+\frac{\kappa^2}{r_0^2}-\omega^2}}+\bar{ D}_{1,2}e^{-r\sqrt{m^2+\frac{\kappa^2}{r_0^2}-\omega^2}}~,
\end{equation} 
and the flux~(\ref{flux2}) at the asymptotic region is given by
\begin{equation}\label{fluxdinfinity}\
\mathcal{F}_{asymp}= \left|\bar{ C}_1\right|^2+\left|\bar{ D}_1\right|^2-\left|\bar{ C}_2\right|^2-\left|\bar{ D}_2\right|^2~,
\end{equation}
up to an irrelevant factor from the angular part of the solution, for $m^2+\frac{\kappa^2}{r_0^2}-\omega^2<0$. On the other hand,  by replacing the Kummer's formula (\ref{KE}), in (\ref{Rhorizon}) and (\ref{psi2horizon}), 
and by using Eq. (\ref{flux2}) we obtain the flux
\begin{equation}\label{fluxdinfinity}
\mathcal{F}_{asymp}= \left|A_1\right|^2+\left|A_2\right|^2-\left|B_1\right|^2-\left|B_2\right|^2~.
\end{equation}
where,
\begin{eqnarray}
\nonumber \bar{ A}_1  &=&\bar{ C}_1\frac{\Gamma \left( c\right) \Gamma \left(
c-a-b\right) }{\Gamma \left( c-a\right) \Gamma \left( c-b\right) }~,\\
\nonumber \bar{ A}_2  &=&\bar{ C}_1\frac{\Gamma \left( c\right) \Gamma \left(a+b-c\right) }{\Gamma \left(a\right) \Gamma \left(b\right) }~,\\
\nonumber \bar{ B}_1  &=&-\bar{ C}_1(m+\frac{i\kappa}{r_0})r_0\frac{\Gamma \left( c\right) \Gamma \left(c-a-b\right) }{\Gamma \left(c+1-a\right) \Gamma \left(c-b\right) }~,\\
\bar{ B}_2  &=&-\bar{ C}_1(m+\frac{i\kappa}{r_0})r_0\frac{\Gamma \left( c\right) \Gamma \left(a+b-c\right) }{\Gamma \left(a\right) \Gamma \left(b+1\right) }~.
\end{eqnarray}
Therefore, the reflection and transmission coefficients
 are given by
\begin{equation}
R=\frac{|\bar{ B}_1|^{2}+|\bar{ B}_2|^{2}}{|\bar{ A}_1|^{2}+|\bar{ A}_2|^{2}}
~,\label{coef1}
\end{equation}
\begin{equation}
T=\frac{|\bar{ C}_1|^{2}}{|\bar{ A}_1|^{2}+|\bar{ A}_2|^{2}}
~,\label{coef2}
\end{equation}
and the absorption cross section $\sigma_{abs}$, becomes
\begin{equation}\label{absorptioncrosssection}\
\sigma_{abs}=\frac{1}{\omega}\frac{|\bar{ C}_1|^{2}}{|\bar{ A}_1|^{2}+|\bar{ A}_2|^{2}}~.
\end{equation}
Now, as in Sec. \ref{coeff}, we will carry out a  numerical analysis of the reflection coefficient~(\ref{coef1}), transmission coefficient~(\ref{coef2}), and absorption cross section~(\ref{absorptioncrosssection}) for  five-dimensional dilatonic black holes,
and we plot the reflection and transmission
coefficient, and the absorption cross section in Figs.~(\ref{Coeff5dk0}, \ref{Coeff5dk3}), for fermionic fields with $m=1$,  $r_0=1$, and $l=0, 1$, respectively. 
Essentially, we have found the same behavior that two-dimensional dilatonic black holes for the coefficients, the reflection coefficient is one in the low frequency limit and for high frequency limit this coefficient is null, being the behavior of the transmission coefficient opposite, and $R+T=1$ occurs in all
cases. 
The absorption cross section is null in the low and high-frequency limit, but there is a range of frequencies for which the absorption cross section is not null. 
Then, in Figs.~(\ref{ACSr0k0}, \ref{ACSr0k3}), we show the variation of the absorption cross section as a function of the horizon radius $r_0$ for $l=0$ and $l=1$ respectively, in this sense the absorption cross section increase if $r_0$ increase. However, beyond a certain value of the horizon $r_0$ the absorption cross section for five-dimensional dilatonic black holes is constant. Besides, we observe that the absorption cross section decreases for higher angular momentum Fig. (\ref{Finall}). Furthermore, we observe that the absorption cross section decreases when the mass of the fermionic field increases Fig. (\ref{Finalm}).

\section{Conclusions}
\label{remarks}
The greybody factor for scalar and fermionic field perturbations in the background of black holes has received great attention. In this context, it was shown that for all spherically symmetric black holes the low energy cross section for massless minimally-coupled scalar fields is always the area of the horizon, where the contribution to the absorption cross section comes from the mode with lowest angular momentum \cite{StarobinskyII, Starobinsky, Das:1996we}. However, for asymptotically AdS and Lifshitz black hole, was observed  
that at low frequency limit there is a range of modes with highest angular momentum, which contribute to the absorption cross section apart of the mode with lowest angular momentum \cite{Gonzalez:2010ht, Gonzalez:2010vv, Gonzalez:2011du,Gonzalez:2012xc}. Also, was observed that the absorption cross section for the three dimensional warped AdS black hole is larger than the area even if the $s$-wave limit is considered, \cite{Oh:2009if}, and recently was found that  
the  zero-angular-momentum greybody factors for non-minimally coupled scalar fields in four-dimensional Schwarzschild-de Sitter spacetime tends to zero in the zero-frequency limit \cite{Crispino:2013pya}. On the other hand, for fermionic fields it was shown that the absorption probability for bulk massive Dirac fermions in higher-dimensional Schwarzschild black hole 
increases with the dimensionality of the spacetime and 
decreases as the angular momentum increases. For this spacetime it was also revealed that the absorption probability depended on mass of the emitted field, that is, the absorption probability decreases or increases depending on the range of energy when the mass of the field increases. Also, was observed that the absorption probability increases for higher radius of the event horizon \cite{Rogatko:2009jp}.

In this work we have studied fermionic perturbations in the background of two and five-dimensional dilatonic black holes, and we have computed the reflection and transmission coefficients, and the absorption cross section, and we have shown numerically that the absorption cross section vanishes at the low and high frequency limit in both cases. Therefore, a wave emitted from the horizon, with low or high frequency, does not reach infinity and is totally reflected, due to the fraction of particles  penetrating the potential barrier vanishes. However, we have shown that there is a range of frequencies where the absorption cross section is not null. The reflection coefficient is one in the low frequency limit and for high frequency limit this coefficient is null, being the behavior of the transmission coefficient opposite, with $R+T=1$. Also, for five-dimensional dilatonic black holes we have shown that the absorption cross section increases if the horizon $r_0$ increases; however, beyond a certain value of the horizon $r_0$ the absorption cross section is constant. It is worth to mention that these results, greybody factors, are consistent with other geometries of dilatonic black holes \cite{Kim:1995hy, Abedi:2013xua}. Besides, we have shown that the absorption cross section decreases for higher angular momentum, and it decreases when the mass of the fermionic field increases.

It is worth to mention that the Dirac equation may be written by making use of the properties of the Dirac operator under conformal transformations \cite{Gibbons:2008rs}. In this case the fermionic field  
\begin{equation}
\psi =\frac{1}{f^{\frac{1}{4}}r_0^{\frac{3}{2}}}e^{-i\omega t}\left( 
\begin{array}{c}
\bar{\psi _{1}} \\ 
\bar{\psi _{2}}%
\end{array}%
\right)\otimes \eta~,
\end{equation}%
allows to reduce the Dirac equation to the following system of differential equations  
\begin{eqnarray}\label{system1}
\sqrt{f}\partial _{r}\bar{\psi _{1}}+\frac{i\omega}{\sqrt{f}}\bar{\psi _{1}}+(m-\frac{i\kappa}{r_0})\bar{\psi _{2}}&=&0~\notag \\
\sqrt{f}\partial _{r}\bar{\psi _{2}}-\frac{i\omega}{\sqrt{f}}\bar{\psi _{2}}+(m+\frac{i\kappa}{r_0})\bar{\psi _{1}}&=&0~.
\end{eqnarray}%
One advantage of this method is to obtain more simple equations, however the physic results, that is, the absorption cross section and the coefficients are the same. 


\section*{Acknowledgments}

We thank to Joel Saavedra for useful comments. This work was funded by Comisi{\'o}n Nacional de  Investigaci{\'o}n Cient{\'i}fica y Tecnol{\'o}gica through FONDECYT Grant 11121148 (Y.V.). P. A. G. acknowledges the hospitality of the Universidad de La Serena where part of this work was undertaken and R. B. acknowledges the hospitality of the Universidad Diego Portales where part of this work was undertaken. P. A. G. and Y. V. acknowledge the hospitality of the National Technical University of Athens.

\end{document}